# Patterning Nanoroads and Quantum Dots on Fluorinated Graphene[*]

Morgana A. Ribas[1], Abhishek K. Singh[1,2], Pavel B. Sorokin[1], and Boris I. Yakobson[1#]

[1] Department of Mechanical Engineering and Materials Science and Department of Chemistry, Rice University, Houston, Texas 77005, USA

[2] Materials Research Centre, Indian Institute of Science, Bangalore 560012, India

[#] Corresponding author: biy@rice.edu

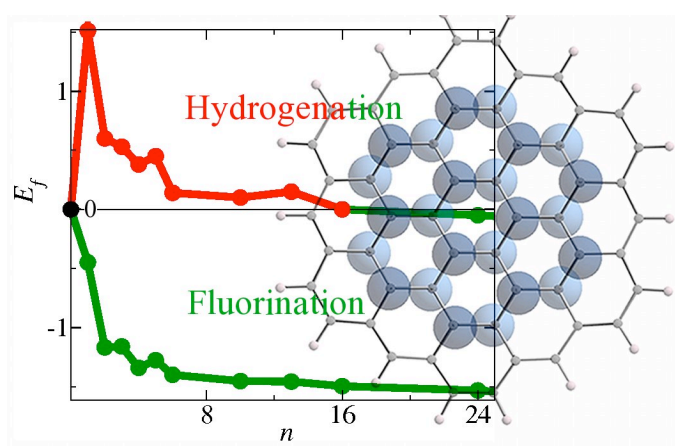

Using ab initio methods we have investigated the fluorination of graphene and find that different stoichiometric phases can be formed without a nucleation barrier, with the complete "2D-Teflon" CF phase being thermodynamically most stable. The fluorinated graphene is an insulator and turns out to be a perfect matrix-host for patterning nanoroads and quantum dots of pristine graphene. The electronic and magnetic properties of the nanoroads can be tuned by varying the edge orientation and width. The energy gaps between the highest occupied and lowest unoccupied molecular orbitals (HOMO–LUMO) of quantum dots are size-dependent and show a confinement typical of Dirac fermions. Furthermore, we study the effect of different basic coverage of F on graphene (with stoichiometries CF and $C_4F$) on the band gaps, and show the suitability of these materials to host quantum dots of graphene with unique electronic properties.

# 1. Introduction

After the first experimental evidence of graphene [1], research on its properties and applications has continued to grow with unprecedented pace. However, a great deal remains to be done to fully incorporate graphene's unique properties into electronic devices. The rapid advances in fabrication methods [2–5] have now made it possible to produce graphene on a large scale. The major obstacle to its application in electronic devices is the lack of a consistent method to open the zero band gap of graphene in a controlled fashion.

The use of graphene nanoribbons has been proposed as a way to tune graphene's electronic properties [6–9]. Depending upon their width and edge orientation, graphene nanoribbons can be either metallic or semiconducting [10–14]. Though there has been intensive research on nanoribbon fabrication, it is still difficult to obtain ribbons with well-defined edges (e.g., by chemical methods [9, 15, 16]) or to scale up its production (e.g., by carbon nanotube unzipping [17, 18]). Even a bottom-up approach [19] does not solve the challenge of assembling graphene nanoribbons into functional devices.

A recently explored alternative way to tune the band gap is to use a fully hydrogenated graphene [20] (also known as graphane [21]) as a matrix-host in which graphene nanoroads [22] or quantum dots [23] are patterned. As is the case for graphene nanoribbons, the electronic properties of such roads and dots are width- and orientation-dependent. However, their biggest advantage over nanoribbons is that both semiconducting and metallic elements can be patterned and interconnected on the same graphene sheet, without compromising its mechanical integrity. This new approach has already been attempted experimentally [24] and has potential to quickly rival graphene ribbons.

The nanoroads and quantum dots can be patterned on any insulating functionalized graphene, e.g., graphane. The formation energy of graphane lies within a favorable and reversible range [25]. This reversibility is very important for applications such as hydrogen storage materials, and has been experimentally observed [26, 27]. However, a nucleation barrier exists in the initial steps of hydrogenation of graphene and it is desirable to find a similar or even more favorable element which can transform graphene into a semiconducting material.

Here we explore the functionalization of graphene by fluorine, which results in compositions similar to Teflon $-(CF_2)_n-$, which in a 2-dimensional incarnation corresponds to $-(CF)_n-$. The biggest advantage of fluorination of graphene comes from the shear availability of experimental and theoretical research that has been done on fluorinated graphite [28–39] and fluorinated carbon nanotubes [40–45]. This knowledge base can help devise ways of controllable fluorination of graphene. Depending on the experimental conditions and reactant gases, different stoichiometries (e.g., $(CF)_n$ [46], $(C_2F)_n$ [29], and $(C_4F)_n$ [47]) can be obtained. Therefore, unlike hydrogenation, fluorination can offer several promising functionalized phases to serve as host materials for graphene nanoroads and quantum dots. Thus, the wide range of possible chemical reactions involving fluorinated graphitic materials, in combination with possibility of tunable electronic properties of patterned graphene structures, opens the door to a range of exciting applications.

Using ab initio methods, we have carried out a comparative study of the formation of CF, $C_2F$, and $C_4F$ (shown in Fig. 1) by chemisorption of F atoms on graphene. Based on the formation energies, CF is the most favorable, and its formation does not have a nucleation barrier, in contrast to the barrier observed for the formation of graphane. We further show that nanoroads and quantum dots of graphene can be patterned on these substrates; these exhibit tunable electronic and magnetic properties, and offer a variety of applications.

# 2. Calculation methods

The calculations were carried out using ab initio density functional theory (DFT), as implemented in the Vienna ab initio package simulation (VASP) [48, 49]. Spin polarized calculations were performed using the projected augmented wave (PAW) method [50, 51] and Perdew–Burke–Ernzerhof (PBE) [52] approximation for the exchange and correlation, with plane wave kinetic energy cutoff of 400 eV. Periodic boundary conditions were used, and the system was considered optimized when the residual forces were less than 0.005 eV/Å. The unit cells lengths of $C_xF_y$ and nanoroads were fully optimized. The Brillouin zone integrations were carried out using a 15 x 15 x 1 Monkhorst–Pack $k$-grid for the $C_xF_y$, 1 x 1 x 5 for the nanoroads, and at the Γ point for the clusters.

For the quantum dots, the total energies for $n$ = 1–384 were calculated using a density functional based tight-binding method (DF–TB) with the corresponding Slater–Koster parameters [53], as implemented in the DFTB+ code [54]. Initially we tested several non-equivalent geometries for the smallest dots ($n$ < 7) and observed that the lowest energy structures were generally derived from $n$–1 structures. The DF–TB results were tested by comparing the formation energies of the smaller dots ($n$ = 1–24) on a finite fully fluorinated graphene cluster ($C_{54}F_{72}$) by DFT calculations (as described above). To mimic an infinite $sp^3$ CF, the edge C atoms of the cluster were passivated with two F atoms and kept constrained during the simulation of the dots. In all calculations sufficient vacuum space was kept between the periodic images to avoid spurious interactions.

## 3. Results and discussion

### 3.1 Structure and electronic properties

The structure of the fluorinated graphite, based on X-ray diffraction results, has long been believed to consist of trans-linked cyclohexane chairs of fluoridated $sp^3$ carbon [28, 29, 36, 37]. Such a structure was later confirmed by DFT calculations [55], and it is the one used here to represent CF (Fig. 1(a)). There are two possible stacking sequences for $(C_2F)_n$: AB/A′B′ and AA′/AA′ [29, 33], where the prime and slash indicate a mirror symmetry and the presence of covalently bonded fluorine atoms, respectively. Both AB (Fig. 1(b)), and AA′ (not shown) are analyzed here. Furthermore, we examine structures of single-sided (Fig. 1(c)) and double-sided fluorinated $C_4F$ [38, 47].

Considering the fluorination to occur through the simple reaction $x$ C + $y/2$ F$_2$ → $C_xF_y$, we calculate the formation energies ($E_f$) of these four structures (Table 1), where $E_f = (E_{C_xF_y} - xE_C - y/2\,E_{F_2})/y$, $E_{C_xF_y}$ is the energy of $C_xF_y$, $E_C$ and $E_{F_2}$ are the energies of a C atom on graphene and F$_2$, respectively, and $x$ and $y$ are the number of C and F atoms, respectively. The formation of CF is more favorable than $C_2F$ and $C_4F$. The formation energy decreases with increasing F coverage. For $C_2F$, AB stacking is energetically more favorable than AA′. The formation energies for $C_4F$ with single- and double-sided fluorination are very similar, within the error of the calculation.

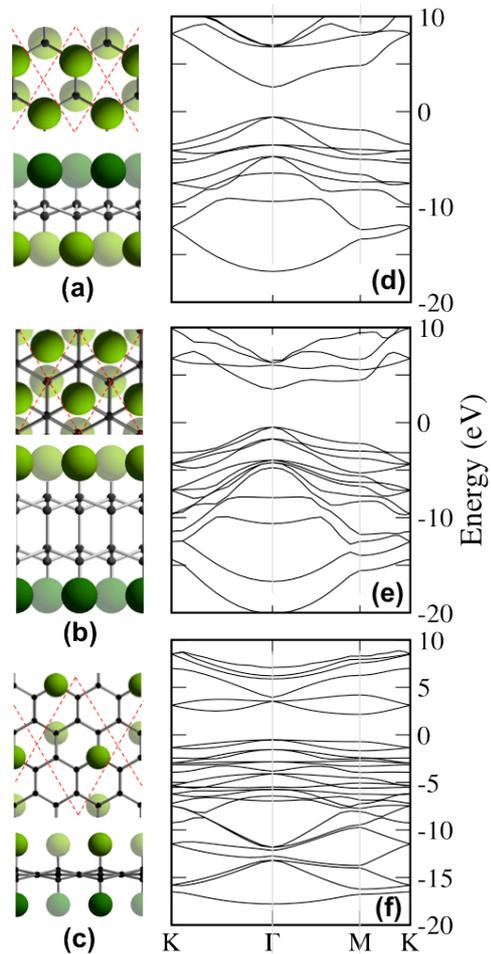

**Figure 1** The atomic structures (darker atoms are closer; red dashed lines mark the unit cells) of (a) CF, (b) $C_2F$ for AB stacking, and (c) $C_4F$ for double-sided fluorination, and (d)–(f) the corresponding electronic band structures. CF and $C_2F$ AB have a direct band gap at the Γ-point, 3.12 eV and 3.99 eV, respectively, whereas $C_4F$ has an indirect band gap of 2.94 eV.

The length of the C–F bond in molecular species is 1.47 Å, and the C-F bond strength is partially attributed to its highly polarized nature. As a result, fluorocarbons have been widely explored in organic chemistry for a variety of applications [32, 56–58]. The calculated C–F bond lengths in CF and $C_2F$ are 1.38 Å in each case and agree well with both experimental (1.41 Å) [33, 37] and theoretical (1.37 Å) [55] values. $C_4F$ shows a much longer C–F bond length (1.45 Å for single-sided fluorination), closer to the C–F bond in molecular species. The computed C–C bond lengths in CF are also in good agreement with both experimental (1.53 Å) [33] and theoretical (1.55 Å) [55, 59] values. The resulting lattice mismatch with graphene ($d_0$ = 2.47 Å) increases with increasing fluorine content. CF has the largest lattice mismatch, 5.7 %, followed by $C_2F$, 3.2 %, and $C_4F$, between 0.4 % (for single-sided coverage) and 0.8 % (for double-sided coverage).

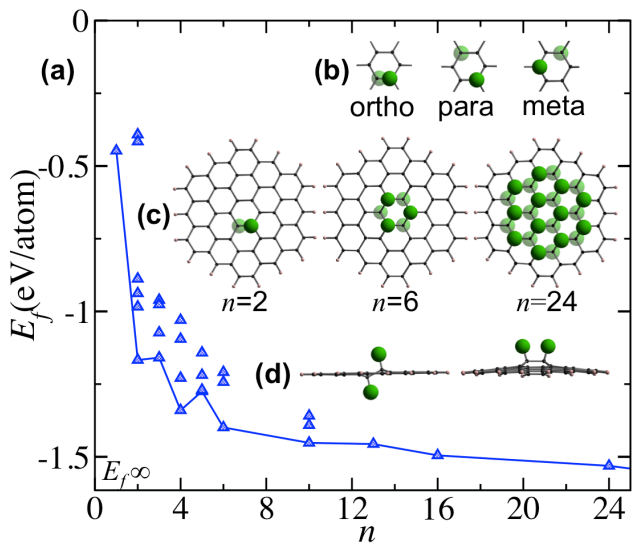

**Figure 2** Chemisorption behavior of F atoms on graphene. (a) The formation energy $E_f$ is negative and further decreases with the number of F atoms $n$, without a nucleation barrier, in contrast to what is observed for hydrogenation [25]. From the initial six positions for the F atom to adsorb (b), an ortho opposite-sided fluorination is lowest in energy, which ultimately results in a preference for aromatic "magic" cluster structures, as illustrated in (c). (d) Example of the lattice strain due to change in hybridization.

Graphene's gapless electronic structure changes completely after fluorination. A finite gap appears in the electronic band structure of CF, $C_2F$, and $C_4F$ (Figs. 1(d)–1(f), respectively), transforming them into wide band gap semiconductors. The electronic band structure of CF shows a 3.12 eV direct band gap at the Γ point, agreeing well with previously reported values [47, 55, 60]. The band gaps of $C_2F$ for both stacking sequences are very similar, which is as expected due to the similarities in their structures. Recent experiments on graphene fluorination yielded an optically transparent $C_4F$ with a 2.93 eV calculated band gap [47]. Here, the calculated band gap for the single-sided fluorinated $C_4F$ is 2.93 eV, a little larger than for the double-sided fluorinated $C_4F$, 2.68 eV.

### 3.2 Formation of fluorinated graphene

The initial steps of graphene fluorination can be studied by incrementally adding $n$ F atoms to different positions of a $C_{54}H_{18}$ cluster (with hydrogenated edges) and calculating the formation energy, $E_f(n) = E_{C_{54}H_{18}F_n} - 1/2nE_{F_2} - E_{C_{54}H_{18}}$. In an aromatic system the π-electrons form pairs between C atoms from different subgroups, starred and un-starred [61]. The adsorption of an odd number of F atoms leaves one unpaired π-electron in the aromatic system. This is exemplified by $E_f$ of a single F atom, which is –0.45 eV/F lower than $1/2E_{F_2}$ (the reference zero on the graphic in Fig. 2(a)). Notably, F attachment is immediately exothermic, in contrast to hydrogenation where H binding is initially an endothermic process (~1.5 eV, relative to molecular $H_2$), causing significant nucleation barrier to the formation of the graphane phases.

Thus, graphene clusters with an odd number of π-electrons are energetically unfavorable and have higher $E_f$, as shown in Fig. 2(a) for $n$ = 1, 3, and 5. For a pair of fluorine atoms there are six different positions (single- and double-sided fluorination, on ortho, meta, and para sites) where they can be added on a ring (Fig. 2(b)). The meta configuration yields the highest $E_f$ because the F atoms bind to C atoms from the same subgroup, therefore creating two unfavorable radicals. Placing the second F atom on

**Table 1** Formation energy ($E_f$), band gap ($E_g$), equilibrium lattice parameter ($d_0$), and bond lengths of the fluorinated graphene at different coverage.

|  | $E_f$, eV/atom | $E_g$, eV | C–F, Å | C–C, Å | $d_0$, Å |
|---|---|---|---|---|---|
| CF | −1.615 | 3.12 | 1.38 | 1.58 | 2.61 |
| $C_2F$ AB | −1.508 | 3.99 | 1.38 | 1.56 | 2.55 |
| $C_2F$ AA' | −1.468 | 3.97 | 1.38 | 1.56 | 2.55 |
| $C_4F$ single-sided | −1.100 | 2.93 | 1.45 | 1.51[a]/1.40[b] | 2.48 |
| $C_4F$ double-sided | −1.095 | 2.68 | 1.48 | 1.50[a]/1.40[b] | 2.49 |

Bond length between [a]$sp^3$ and $sp^2$ C atoms, and [b]two $sp^2$ C atoms.

the opposite side in the ortho-configuration results in a lower $E_f$; the value is 1.17 eV/atom lower than $1/2 E_{F_2}$, which is as expected since the radical on the adjacent C atom is completely passivated.

Adsorption of an F atom transforms the hybridization of the host C atom from $sp^2$ to $sp^3$. As a result of this, the bond between the $sp^3$ and $sp^2$ carbon atoms becomes elongated, which leads to strain in the structure and an out of plain buckling of the fluorinated C atom. By adding a second F atom on the opposite side, the induced strains compensate one another further lowering the total energy (Fig. 2(c)). The $E_f$ decreases with increasing F content and approaches asymptotically to $E_f(\infty)$, which is equal to the formation energy of a fully fluorinated infinite graphene sheet.

According to the definition of the formation energy, fluorination of the graphene is favorable when $E_f(n) < 1/2 E_{F_2}$. In the case of the analogous structure of graphane, the formation energy is lower than $1/2 E_{H_2}$ only after the formation of a stable nucleus formed by adsorption of 24 hydrogen atoms [25]. Since the C–F bond is stronger than the C–H bond and the F–F bond in $F_2$ is much weaker than the H–H bond in $H_2$, fluorination of the graphene is more favorable than its hydrogenation. In fact, we observe that $E_f(n)$ is always lower than $1/2 E_{F_2}$. Therefore, unlike graphane, fluorinated graphene does not have a nucleation barrier to its formation and should be more stable and easily obtained than graphane.

### 3.3 Graphene nanoroads on CF

In order to explore the possibility of combining metallic and semiconducting properties on the same sheet, we investigated the patterning of nanoroads (NR′) and quantum dots on fluorinated graphene (FG). Graphene nanoroads were formed by removing F atoms from a fully fluorinated graphene either along armchair (AC) dimers ($N_{ac}$) or zigzag (ZZ) chains ($N_{zz}$) to form pristine graphene roads (Figs. 3(a) and 3(b)). After geometrical optimization, the AC nanoroad (AC-FGNR′) remains flat whereas the ZZ nanoroad (ZZ-FGNR′) is tilted. This tilting is a geometrical consequence of the position of the F atoms along the road being alternated in and out of the plane.

The AC-FGNR′ are semiconducting with large band gaps due to quantum confinement (Fig. 3(c)). The band gap behavior can be divided into three hierarchical families for $N_{ac}$ = 3p, 3p+1, and 3p+2, where p is a positive integer. For the AC orientation, the band gaps do not follow a monotonic trend with the width, and instead $\Delta_{N_{ac}=3p+2} < \Delta_{N_{ac}=3p+1} < \Delta_{N_{ac}=3p}$ (except for $N_{ac}$ = 3 and 4). This trend is different from that observed for armchair nanoribbons [11], but is similar to that for graphane nanoroads [22]. The band gap energies are not affected by increasing the distance between the roads by adding additional chains of fluorinated graphene. Thus the AC-FGNR′ are also well isolated from their lateral periodic images.

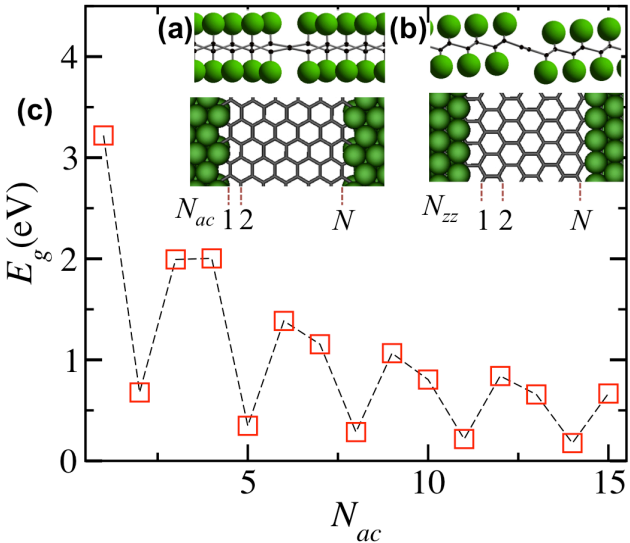

**Figure 3** Schematic illustrations of (a) armchair (AC) ($N_{ac}$ is the number of $sp^2$ C dimer lines) and (b) zigzag (ZZ) ($N_{zz}$ is the number of $sp^2$ C chains) roads. (c) For AC-FGNR′, the band gap $E_g$ energy varies with the road width $N_{ac}$.

The electronic properties for the ZZ-FGNR′ depend strongly on the magnetic states of the system. We observe that the antiferromagnetic ZZ-FGNR′ are semiconducting, whereas the ferromagnetic ZZ-FGNR′ are semi-metallic. Very narrow nanoroads on graphane were found to be nonmagnetic, with a very small gap, and two bands near the Fermi level that do not cross. In contrast, narrow ZZ-FGNR′ have larger band gaps, for example $E_g$ is 0.71 eV for $N_{zz}$ = 1 and 0.56 for $N_{zz}$ = 2. This difference may be due to the larger lattice mismatch between the fluorinated and pristine graphene. The states with antiferromagnetic spin distribution are lower in energy than the ferromagnetic, which can be used for spintronic applications.

For ZZ-FGNR′, the band gap energy increases with decreasing width (Fig. 4(a)) due to quantum confinement. We find that the band gap is inversely proportional to the width $N_{zz}$, which is similar to the dependence observed for nanoribbons [11, 62] or graphene with periodically adsorbed hydrogen chains [63], showing the best fit for $E_g(N_{zz}) = 4.47/(5.69 + N_{zz})$ eV. There are only two bands which cross near the Fermi level of the ZZ-FGNR′. We plot their corresponding band-decomposed charge densities in Figs. 4(b) and 4(c) and observe that the π-bands formed by the overlap of $p_z$ orbitals are mostly localized on the C atoms at the $sp^2$–$sp^3$ interface.

### 3.4 Graphene quantum dots on CF and C$_4$F

Experimentally, graphene quantum dots are obtained by cutting tiny pieces of graphene into different shapes; however, the problem of how to tune graphene's band gap still persists. Our approach is to consider quantum dots as small islands of graphene created by the removal of the F atoms from CF and C$_4$F. We study their thermodynamic feasibility by analyzing several possible configurations and explore how quantum confinement affects their electronic properties. The removal of the F atoms yields graphitic islands of connected $n$ $sp^2$ C atoms on a fully fluorinated graphene finite cluster or infinite sheet. For $(C_xF)_n \rightarrow (C_xF)_{N-n}C_n + n/2\,F_2$, the formation energy is calculated as $E_f(n) = [E_{sys}(n) - n\mu_F - N\mu_{C_xF}]/n$, where $E_{sys}$ is the total energy of quantum dots on the fluorinated graphene, and $\mu_F$ (= $1/2 E_{F_2}$) and $\mu_{C_xF}$ are the chemical potentials of fluorine and CF, respectively. The removal of one F atom introduces an additional π radical on the C atoms; this is the inverse of F adsorption on graphene shown above.

Formation of dots with lower $E_f(n)$ is more favorable and, overall, the formation energy of quantum dots decreases with increasing size of the dots. This trend can be observed by both DFT and

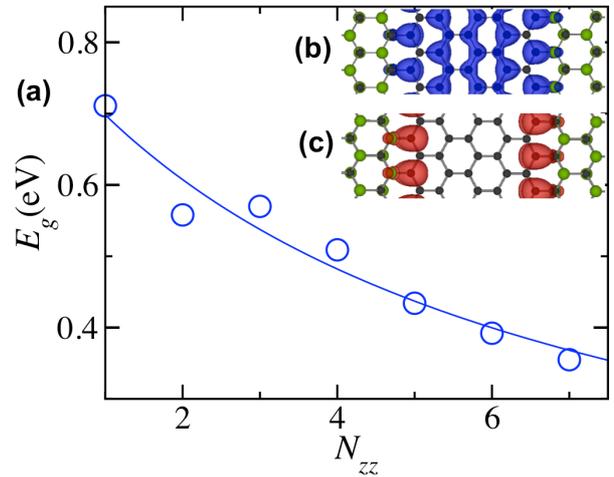

**Figure 4** (a) Plot showing the decrease in band gap with increasing width of the road for ZZ-FGNR′. The band-decomposed charge densities (3 x $10^{-3}$ Å$^{-3}$) for the 5-ZZ-FGNR′, corresponding to the top of the valence band (b) and the bottom of the conduction band (c).

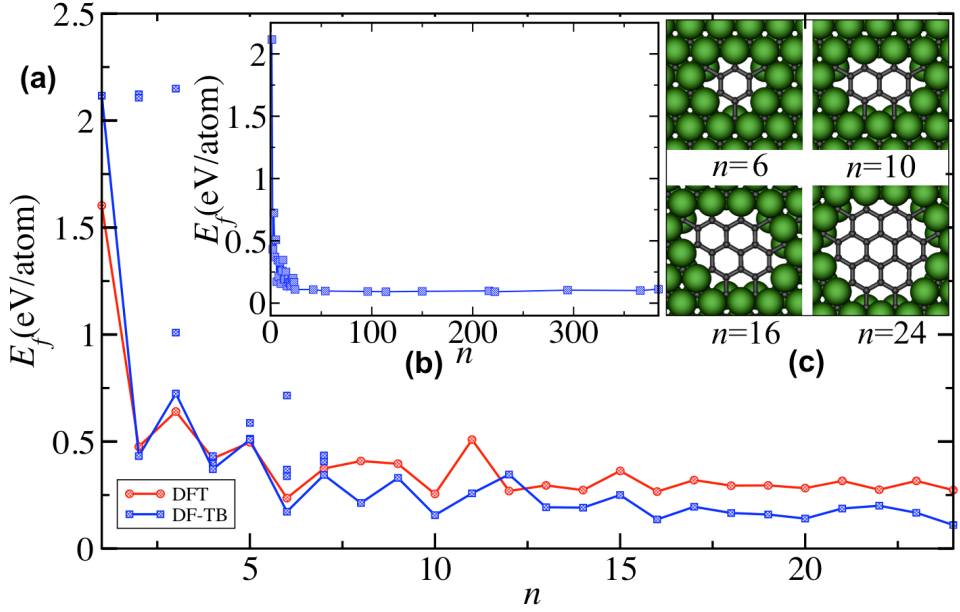

**Figure 5** (a) Formation energy $E_f(n)$ for different sizes of quantum dots calculated by DFT and DF–TB. (b) Overall, $E(n)-E(\infty) \sim const/\sqrt{n}$, using DF–TB. (c) Examples of aromatic quantum dots.

DF–TB for dots carved on CF (Fig. 5(a)). Larger dots can be studied using DF–TB, and upon their inclusion a clear trend, where $E(n) - E(\infty) \sim const/\sqrt{n}$, can be observed (Fig. 5(b)). The quantum dots with the lowest formation energies are those in which the structure restores the aromaticity of the graphene, for example $n$ = 6, 10, 16, and 24 (Fig. 5(c)).

The realization of quantum confinement promises many exciting applications like quantum computing [64, 65], single-electron transistors [66], and optoelectronics. The band gap is important for optical applications, but DFT and DF–TB calculations generally underestimate its value. The DF–TB band gaps at the Γ point for $n$ = 6 (3.28 eV) and 24 (2.44 eV) agree well with the HOMO–LUMO gaps calculated by DFT (3.37 and 2.47 eV, respectively for $n$ = 6 and 24), therefore lending additional credibility to the DF–TB method employed. Although it is computationally prohibitive to calculate the precise band gaps of large systems ($n$ > 24), these will most probably be within the optical range of ~1–3 eV.

There is a reduction in quantum confinement with increasing size of the dots, thus reducing the band gap energy as their electronic properties approach those of graphene. Larger dots can be divided into AC and ZZ edges (examples shown in the inset in Fig. 6). Interestingly, plots of their band gaps show that dots with AC edges have larger band gaps than those with ZZ edges (Fig. 6). A least squares fit also shows the different trends for AC and ZZ edges; $E_g(n) = 14.1n^{-1/2+0.01}$ eV for AC edges and $E_g(n) = 19.4n^{-1/2-0.14}$ eV for ZZ edges. Different from conventional quantum dots, which follow a ~$1/R^2$ dependence, the observed trend is closer in behavior

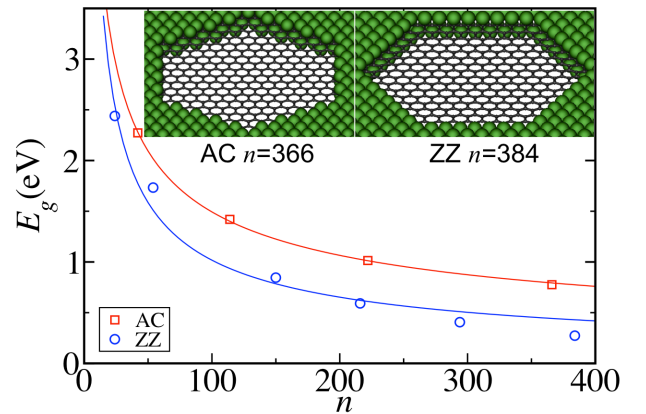

**Figure 6** Plots showing how the energy of the band gap for dots with AC and ZZ edges decreases with their size. Inset: Configurations of the largest optimized quantum dots.

to the $1/R$ confinement of Dirac fermions [67, 68], where $R \sim \sqrt{n}$.

Recently synthesized fluorinated graphene films were found to be optically transparent at the $C_4F$ saturation level [47]. Next we study three examples of 2D quantum dot arrays, as this is the way they are usually assembled for optical applications. We carved a graphene patch as coronene on a 7 x 7 hexagonal supercell of CF ($n = 24$) and on a 4 x 4 hexagonal supercell of $C_4F$ ($n = 6$). All dots were separated by at least 10 Å.

The band-decomposed electron densities of the dots show atom-like states for CF and $C_4F$. The band gaps of the arrays are very similar to the ones obtained for the isolated dots, 2.50 eV for CF and 0.88 eV for $C_4F$. The nearly dispersionless bands for the dot on CF at the top of the valence band and at the bottom of the conduction band (Fig. 7(a)) show a very good quantum confinement, with just a very small charge density "leakage" into the fluorinated graphene. In fact, the electronic state of the quantum dots on CF seems to be somehow more confined than on similar dots on graphane [23] due to the insulator character, which comes from the charge transfer from C to F in CF. The array of $C_4F$ also shows localized charge density states (Fig. 7(b)) with a non-hexagonal shape. The difference between the shape of the dots results from the more "open" structure of $C_4F$. Thus both CF and $C_4F$ can be used for patterning quantum dots and, most importantly, the differences in fluorine coverage on graphene can be also used to tune the band gap.

## 4. Conclusions

We have investigated a way to alter gapless graphene by its patterned fluorination. Such fluorination results in wide band gap semiconductors (CF, $C_2F$, and $C_4F$), where higher F coverage is favored. We find fluorination of graphene to be different from its hydrogenation, as it occurs without a nucleation barrier owing to the higher affinity of F towards C. Furthermore, the suitability of CF as a host material for graphene nanoroads and quantum dots has been demonstrated. We find that nanoroads and quantum dots exhibit orientation-, width-, and F coverage-dependent electronic properties. Fluorinated graphene nanoroads with AC orientation are semiconducting with large band gaps, following a non-monotonic variation. The nanoroads with ZZ edge are semiconducting or semi-metallic according to their spin orientation; antiferromagnetic or ferromagnetic, respectively. The band gaps in ZZ roads vary as $\sim 1/N_{zz}$. The formation energy of the quantum dots depends on their size as $E(n) - E(\infty) \sim const/\sqrt{n}$. The band gaps of the larger quantum dots follow a $1/R$ trend similar to the confinement of Dirac fermions, where $R \sim \sqrt{n}$. The band-decomposed electron densities of 2D quantum dot arrays in CF and $C_4F$ show atom-like states with very good electron confinement.

Current developments in techniques for the functionalization of graphene should make the patterns shown here an experimental possibility in near future. These patterns can serve as building blocks, which can be formed on different phases of fluorinated graphene (as shown here for CF and $C_4F$), opening the door to a range exciting of applications.

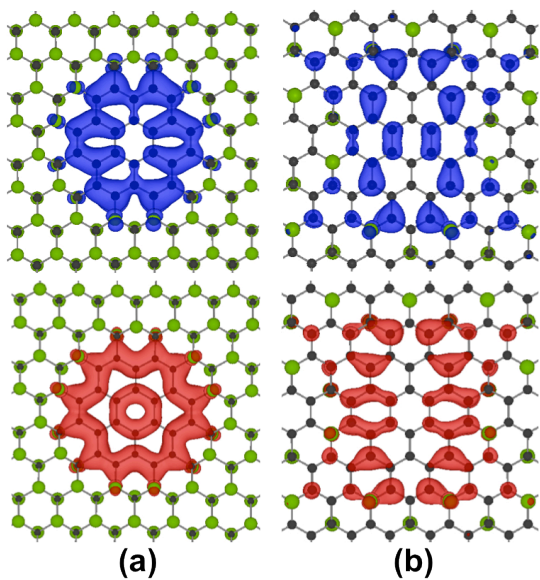

**Figure 7** Isosurfaces of band-decomposed charge densities (1.5 x $10^{-4}$ Å$^{-3}$) at the top of the valence band (upper figures in blue) and at the bottom of the conduction band (lower figures in red) for 2D quantum dot arrays on (a) CF for $n = 24$, and $C_4F$ for (b) $n = 6$.

We realize that, as in the case of any chemical "attack" [69], fluorination can cause some defects and the degree of such damage will depend on the specific conditions (source of F, temperature, etc.). Another uncertainty is due to the possible coexistence of different configurations (chair, boat, etc.), which is not within the scope of this study. The emphasis here is simple: Most of these fluorinated graphene phases have a sizable gap, and therefore can serve as a host matrix for confined nearly-metallic domains of pristine carbon. It will be important to further explore the interface and to learn how to avoid possible frustration [70] which may destroy the interfaces and the clear picture of confinement, calling for further careful studies. One can speculate that possible experimental approaches fall into two classes: Either one can mask certain areas prior to exposing them to fluorination, or start from fully fluorinated graphene and then attempt a local removal of F (say, with focused ion beam of not too high an energy).

## Acknowledgements

This work was supported by the Office of Naval Research (MURI project). The computations were performed on Kraken, at the National Institute for Computational Sciences, through allocation No. TG-DMR100029. M.A.R. is partially supported by a fellowship from the Roberto Rocca Educational Program.

*Note added in proof*: After the completion of this work, an experimental evidence of stoichiometric fluorinated graphene was reported [71]. The authors report that the CF phase is indeed more stable that graphane, further corroborating the present work.